\documentclass[prl,twocolumn,tightenlines,superscriptaddress,nofootinbib,showpacs]{revtex4}
\usepackage{amssymb,latexsym}
\usepackage{amsmath,amsbsy}
\usepackage{epsfig,bm}
\usepackage{graphicx,comment}
\usepackage{epsf}
\usepackage{amscd}
\usepackage{graphicx,comment}

\DeclareMathOperator{\str}{str}
\DeclareMathOperator{\diag}{diag}

\def\a{{\alpha}}

\def\d{{\delta}}

\def\ol#1{{\overline{#1}}}

\def\Dslash{D\hskip-0.65em /}

\def\CPT{{$\chi$PT}}

\def\PQCPT{{PQ$\chi$PT}}

\def\cL{{\mathcal L}}

\def\cM{{\mathcal M}}

\begin{document}

\preprint{UMD-40762-413}

\author{F.-J.~Jiang}
\email[]{fjjiang@itp.unibe.ch}
\affiliation{%
Institute for Theoretical Physics,
Bern University,
Sidlerstrasse 5,
CH-3012 Bern, 
Switzerland
}

\author{B.~C.~Tiburzi}
\email[]{bctiburz@umd.edu}
\affiliation{%
Maryland Center for Fundamental Physics, 
Department of Physics, 
University of Maryland, 
College Park,  
MD 20742-4111, 
USA
}


\title{Flavor Twisted Boundary Conditions in the Breit Frame}

\begin{abstract}
We use a generalization of chiral perturbation theory to account for
the effects of flavor twisted boundary conditions in the Breit frame. 
The relevant framework for two light flavors is an 
$SU(6|4)$
partially quenched theory, where the extra valence 
quarks differ only by their boundary conditions.
Focusing on the pion electromagnetic form factor, 
finite volume corrections are calculated at next-to-leading
order in the chiral expansion and are estimated to be small on current lattices.
\end{abstract}


\maketitle




\noindent
{\bf Twisted Boundary Conditions and the Breit Frame}.
Simulations of QCD on Euclidean spacetime 
lattices are making progress towards quantitative 
understanding of strong interactions~\cite{DeGrand:2006aa}.
A source of systematic error in these calculations
is the finite lattice volume. 
While observables generally depend upon the lattice
volume, there is a potentially more serious effect:
periodic boundary conditions on quark fields 
limit the available momentum modes to integer multiples of 
$ 2 \pi / L$, 
where 
$L$
is the size of the lattice. 
This presents difficulty for the study of
hadronic properties at small momentum.
Periodic boundary conditions, however, are often chosen for convenience;
and, 
a more general class of boundary conditions, 
twisted boundary conditions (TwBCs), 
see e.g.~\cite{Zinn-Justin:1996cy}, 
are possible.
A TwBC by an arbitrary twist angle
$\theta$
on the matter field 
$\psi$, 
has the form:
$\psi(x+L) = e^{ i \theta} \psi(x)$,
and consequently the matter field has kinematic momentum
$p = ( 2 \pi n + \theta) / L$,
where 
$n$ 
is an integer.
The ability to produce continuous hadronic momentum has
motivated the study of TwBCs in lattice QCD~%
\cite{Guagnelli:2003hw,Bedaque:2004kc,deDivitiis:2004kq,Sachrajda:2004mi,Bedaque:2004ax,Tiburzi:2005hg,Flynn:2005in,Guadagnoli:2005be,Aarts:2006wt,Tiburzi:2006px,Jiang:2006gna,Boyle:2007wg,Simula:2007fa,Boyle:2008yd}.

Producing continuous 
hadronic momenta via TwBCs
does not necessarily give one the ability to 
probe amplitudes at continuous momentum transfer. 
In fact, these techniques are of no avail for  flavor singlet
form factors which require operator self-contractions. 
In a current self-contraction, 
the boundary condition chosen for the quark in the current
does not affect the momenta of the external states.
Flavor non-singlet form factors, however, can be accessed at continuous
momentum transfer. 
A conceptually clear way to see this is to consider form factors of 
flavor changing currents~\cite{Tiburzi:2005hg}. 
If the current induces a change from a quark flavor satisfying periodic boundary conditions to 
one satisfying TwBCs, continuous momentum transfer, 
$\theta / L$, 
will be induced. 
There is another way to utilize TwBCs for flavor non-singlet currents~\cite{Boyle:2007wg}. 
For this method, imagine that the current strikes a quark of a given twist angle 
$\theta$,
and then produces a different twist angle 
$\theta'$. 
Here the momentum transfer 
$(\theta' - \theta) / L$ 
is induced. 
In the infinite volume limit, these boundary conditions become irrelevant
and the current thus produces momentum transfer by striking a single
(non self-contracted) quark. 
Furthermore one can choose $\theta' = - \theta$ to work in the 
Breit frame, 
 which often simplifies the calculation of form factors. 
At finite volume, however,
the initial- and final-state quarks are actually distinct
because a given field can only have one boundary condition. 
Hence the lattice action must be described by a theory
with an enlarged valence flavor group~\cite{Boyle:2007wg}.
The purpose of this note is to utilize the effective 
field theory for the Breit frame implementation of 
TwBCs, discussed in~\cite{Boyle:2007wg}, 
to compute volume corrections to observables, 
in particular the pion electromagnetic form factor.

\noindent
{\bf Partially Quenched Chiral Perturbation Theory}.
%
%
%
%
To address the long-distance effects of TwBCs in the Breit frame, we must formulate the low-energy effective theory. 
From the discussion in~\cite{Boyle:2007wg}, 
this variant of chiral perturbation theory 
includes additional fictitious valence quarks. 
These fictitious quarks only differ by their boundary conditions.
We can remove the twists by a field 
redefinition in favor of periodic quark fields. 
In terms of these fields, the quark part of the partially twisted, partially quenched 
QCD Lagrangian appears as
\begin{equation} \label{eq:LQCD}
\cL = \ol Q \left( \tilde{\Dslash} + m_Q \right) Q
,\end{equation}
where the vector 
$Q$ 
accommodates ten quark fields, 
$Q = ( u_1, u_2, d_1, d_2, j, l, \tilde{u}_1, \tilde{u}_2, \tilde{d}_1, \tilde{d}_2 )^T$. 
The field 
$Q$ 
transforms under the fundamental representation of the graded group 
$SU(6|4)$. 
The quark mass matrix in the isospin limit%
\footnote{
Although we work with $m_d = m_u$, we have kept the 
flavor group to be $SU(6|4)$ 
[as opposed to $SU(5|3)$]
to allow for strong isospin breaking, $m_d \neq m_u$. 
} 
of the valence and sea sectors
is given by 
$m_Q = \diag (m_u, m_u, m_u, m_u, m_j, m_j, m_u, m_u, m_u, m_u)$. 
Because we implement twisted boundary conditions only in the valence sector, 
we have additionally chosen to work away from the unitary point, where
$m_j = m_u$. 
With this choice, we can track sea-quark effects with ease
and account for partial quenching errors. 
Formulae relevant at the unitary point can easily be recovered 
by taking the limit 
$m_j \to m_u$.

The 
$Q$ 
field is periodic and the effects of partially TwBCs
 have been shuffled into the covariant derivative 
$\tilde{\Dslash}$, 
where they have the form of a uniform 
$U(1)$ 
gauge potential 
$B_\mu$:
$\tilde{D}_\mu = D_\mu + i B_\mu$. 
The quark flavors are charged differently under this
$U(1)$ 
potential, specifically the flavor matrix is given by
$B_\mu =  \diag (B_\mu^{u}, - B_\mu^{u}, -B_\mu^d,  B_\mu^d, \, 0, \, 0,  B_\mu^{u}, - B_\mu^{u}, -B_\mu^d,  B_\mu^d)$.
Because the
$U(1)$ 
field is uniform, the 
$B_\mu$ 
act as flavor dependent field momenta, each having the form 
$B^q_\mu = (\bm{\theta}^q / L, 0)$,
where 
$L$ 
is the spatial length of the lattice and 
$q$ 
is a flavor index. 
Notice there is no fourth component to 
$B_\mu$
because the boundary conditions are only 
spatially twisted. 
The twist angles 
$\bm{\theta}$
can be implemented on the lattice by 
modifying the links,
$U_j(x) \to U_j(x) e^{i \theta^q_j / L}$, 
where 
$j$ 
is a spatial index. 
Notice that the twist angles vanish for the sea-sector, 
this corresponds to the partially twisted scenario
where existing gauge configurations can be 
post-multiplied by the uniform 
$U(1)$ 
gauge potential.
We additionally must choose the ghost 
quarks to be degenerate with their valence 
counterparts, and further to satisfy the same
boundary conditions as their valence counterparts. 
The determinant thus arises only from the sea quarks. 
The graded symmetry of 
$SU(6|4)$ 
hence provides a way to write down a theory
corresponding to partially twisted, partially quenched
QCD, see~\cite{Sharpe:2001fh} for a more rigorous discussion.

The low-energy effective theory of partially quenched QCD
 is partially quenched chiral perturbation theory (PQ\CPT), 
which describes the dynamics of pseudo-Goldstone mesons 
arising from spontaneous chiral symmetry breaking.  
In finite volume, we restrict our analysis to the 
$p$-regime~\cite{Gasser:1987zq}
in which the long-range fluctuations of the Goldstone modes
cannot conspire to restore chiral symmetry. 
The effective theory is written in terms of 
the periodic coset field 
$\Sigma = \exp ( 2 i \Phi / f)$. 
Here the meson fields
$\Phi$ 
are embedded non-linearly,
and in our conventions the parameter 
$f$ 
is the pion decay constant, 
$f \sim 130 \, \texttt{MeV}$.

To obtain the relevant generalization of  PQ\CPT\  for
Eq.~\eqref{eq:LQCD}, 
we take 
$\Sigma$ 
to transform as 
$\Sigma \to L \Sigma R^\dagger$
under a chiral transformation 
$(L,R) \in U(6|4)_L \otimes U(6|4)_R$
and write down the most general
chirally invariant Lagrangian.
Furthermore, we include the uniform gauge
potential by requiring the theory be invariant 
under local 
$U(1)_V$ 
phase rotations, see~\cite{Sachrajda:2004mi}.
In terms of 
$\Sigma$, 
the Lagrangian of PQ\CPT\ is
\begin{equation} \label{eq:L}
\cL 
= 
\frac{f^2}{8} 
\str \left( \tilde{D}_\mu \Sigma \tilde{D}_\mu \Sigma^\dagger \right)
-
\lambda \str \left( m_Q^\dagger \Sigma + \Sigma^\dagger m_Q \right)
+
\mu_0^2 \Phi_0^2
,\end{equation}
where we have written down only the leading order 
terms in an expansion in 
$m_Q$ 
and 
$p^2$, 
where 
$p$ 
is a momentum. 
Here 
$\Phi_0$
is the flavor singlet field
$\Phi_0 = \str ( \Sigma) / \sqrt{2}$, 
and the action of the covariant derivative
$\tilde{D}_\mu$
is specified by
\begin{equation}
\tilde{D}_\mu \Sigma 
=
\partial_\mu \Sigma
+ 
i [ B_\mu, \Sigma]
+ 
i A^a_\mu 
[\ol T {}^a, \Sigma ] 
.\end{equation} 
We have included an isovector
source field 
$A^a_\mu$ 
which will be utilized below. 
The meson modes are contained in $\Phi$,
which is a ten-by-ten matrix of fields.
It has the form
\begin{equation}
\Phi
=
\begin{pmatrix}
M_{vv} & M_{vs} & \chi^\dagger_{gv} \\
M_{sv} & M_{ss} & \chi^\dagger_{gs} \\
\chi_{gv} &  \chi_{gs} & M_{gg} 
\end{pmatrix}
.\end{equation}
The mesons of $M_{vv}$ ($M_{gg}$)  are bosonic and 
are formed from a valence (ghost) quark-antiquark pair. 
These matrices have the form
\begin{widetext}
\begin{equation}
M_{vv}
=
\begin{pmatrix}
\eta^u_{11}  & \eta^u_{12} & \pi^+_{11}  & \pi^+_{12} \\
\eta^u_{21}  & \eta^u_{22} & \pi^+_{21} & \pi^+_{22} \\
\pi^-_{11} & \pi^-_{12} & \eta^d_{11} & \eta^d_{12} \\
\pi^-_{21} & \pi^-_{22} & \eta^d_{21} & \eta^d_{22}
\end{pmatrix}
, \, \,
\text{and}
\quad
M_{gg}
=
\begin{pmatrix}
\tilde{\eta}^u_{11}  & \tilde{\eta}^u_{12} & \tilde{\pi}^+_{11}  & \tilde{\pi}^+_{12} \\
\tilde{\eta}^u_{21}  & \tilde{\eta}^u_{22} & \tilde{\pi}^+_{21} & \tilde{\pi}^+_{22} \\
\tilde{\pi}^-_{11} & \tilde{\pi}^-_{12} & \tilde{\eta}^d_{11} & \tilde{\eta}^d_{12} \\
\tilde{\pi}^-_{21} & \tilde{\pi}^-_{22} & \tilde{\eta}^d_{21} & \tilde{\eta}^d_{22}
\end{pmatrix}
\notag
.\end{equation}
The 
$\eta^q_{ij}$ ($\tilde{\eta}^q_{ij}$)  
mesons have quark content 
$\eta^q_{ij} \sim  q_i  \ol q_j$
($\tilde{\eta}^q_{ij} \sim  \tilde{q}_i  \overline{\tilde{q}}_j$), 
while the 
$\pi^+_{ij}$ 
($\tilde{\pi}^+_{ij}$)
mesons have quark content
$\pi^+_{ij} \sim u_i \ol d_j$ 
($\tilde{\pi}^+_{ij} \sim \tilde{u}_i \overline{\tilde{d}}_j$ ).
The valence-sea (sea-sea) mesons are bosonic and contained in 
$M_{vs}$ 
($M_{ss}$)
as
\begin{equation}
M_{sv}
=
\begin{pmatrix}
\phi_{j u_1}  & \phi_{j u_2} & \phi_{j d_1} & \phi_{j d_2} \\
\phi_{l u_1}  & \phi_{l u_2} & \phi_{l d_1} & \phi_{l d_2}
\end{pmatrix}
, \,\,
\text{and} 
\quad
M_{ss} 
=
\begin{pmatrix}
\eta_j & \pi_{jl} \\
\pi_{lj} & \eta_l
\end{pmatrix}
\notag
.\end{equation}
Mesons contained in $\chi_{gv}$ ($\chi_{gs}$)
are built from ghost quark, valence antiquark (sea antiquark) pairs
and are thus fermionic. These states appear as
\begin{equation}
\chi_{gv}
=
\begin{pmatrix}
\phi_{\tilde{u}_1 u_1} & 
\phi_{\tilde{u}_1 u_2} & 
\phi_{\tilde{u}_1 d_1} & 
\phi_{\tilde{u}_1 d_2} \\ 
\phi_{\tilde{u}_2 u_1} & 
\phi_{\tilde{u}_2 u_2} &
\phi_{\tilde{u}_2 d_1} & 
\phi_{\tilde{u}_2 d_2} \\
\phi_{\tilde{d}_1 u_1} & 
\phi_{\tilde{d}_1 u_2} & 
\phi_{\tilde{d}_1 d_1} & 
\phi_{\tilde{d}_1 d_2} \\
\phi_{\tilde{d}_2 u_1} & 
\phi_{\tilde{d}_2 u_2} & 
\phi_{\tilde{d}_2 d_1} & 
\phi_{\tilde{d}_1 d_2} 
\end{pmatrix}
, \, \, 
\text{and}
\quad
\chi_{gs}
=
\begin{pmatrix}
\phi_{\tilde{u}_1 j} & 
\phi_{\tilde{u}_1 l} \\ 
\phi_{\tilde{u}_2 j} & 
\phi_{\tilde{u}_2 l} \\ 
\phi_{\tilde{d}_1 j} & 
\phi_{\tilde{d}_1 l} \\ 
\phi_{\tilde{d}_2 j} & 
\phi_{\tilde{d}_2 l} 
\end{pmatrix}
\notag
.\end{equation}
%
%
%
\end{widetext}
%
%
%

In writing the PQ\CPT\ Lagrangian
we have kept the flavor singlet field 
$\Phi_0$ as a device.
Expanding $\cL$ to tree level, 
one 
finds 
that mesons with 
quark content $\sim Q \ol Q '$ have 
masses:
$m_{QQ'}^2 
= 
\frac{\lambda}{4 f^2}
\left( m_Q + m_{Q'} \right)$.
The flavor singlet additionally
acquires a mass 
$\mu_0^2$. 
Taking the limit 
$\mu_0 \to \infty$ 
(which is warranted by the strong axial anomaly),
we integrate out the flavor singlet
component. 
The resulting Goldstone manifold is 
$SU(6|4)$, 
but only the propagators of the flavored mesons
have simple forms. The neutral meson propagators
have both flavor connected and disconnected
(hairpin) terms. 
The flavor neutral propagator 
was derived in general for the
$SU(M + N | N)$
group in~\cite{Sharpe:2001fh}
and is identical for our case because
the flavor neutral states are unaffected by twisting. 
We will not need the explicit form
of this propagator in order to 
display our final results.

Further calculation at tree level
shows that mesons with quark content 
$Q \ol Q '$ 
have kinematic momentum
$\bm{P}_{Q Q'}$
given by
$\bm{P}_{Q Q'}
=
\bm{p} + \bm{B}_Q - \bm{B}_{Q'}$.
At finite volume, mass splittings are generated,
and kinematic momenta are renormalized by infrared effects~\cite{Jiang:2006gna}.
These effects were estimated to be on the order of a few percent on current lattices.
As we shall calculate finite volume corrections to 
the pion form factor, which themselves are at the 
percent level, we can safely ignore mass splittings
and momentum renormalization below.



\noindent
{\bf Pion Form Factor in Finite Volume}.
To calculate the pion form factor on the lattice
using TwBCs in the Breit frame, 
two separate current insertions are used. 
In the first (second) insertion, the current
strikes the up (anti-down) quark. 
Matching to this lattice setup, in our theory we have%
\footnote{
There are three other combinations of matrix elements
that lead to the same infinite volume physics. 
This is due to the redundancy of the spectator
quarks, which are evaluated at zero twist angle.
We have verified that each of the four matrix
elements calculated in the effective theory
yield the same answer at infinite and finite volume. 
} 
\begin{eqnarray}
&& 
\langle \pi^+_{21} (\bm{0}) | J_\mu^1 | \pi^+_{11} (\bm{0}) \rangle \Big|_{\bm{\theta}^d = \bm{0}}
+
\langle \pi^+_{22} (\bm{0}) | J_\mu^2 | \pi^+_{21} (\bm{0}) \rangle \Big|_{\bm{\theta}^u = \bm{0}}
\notag \\
&&  \phantom{middlespaces} 
\overset{L \to \infty}{\longrightarrow}
\langle \pi^+(-\bm{p}) | J_\mu | \pi^+(\bm{p}) \rangle
,\end{eqnarray}
with 
$\bm{p} = \bm{\theta} / L$ 
fixed.  The currents 
$J_\mu^{1,2}$ 
correspond to the photon hitting the 
$u$ 
and 
$\ol d$ 
quarks, respectively: 
\begin{eqnarray}
J_\mu^1 &=& q_u \, \ol u_2 \gamma_\mu u_1 \\
J_\mu^2 &=& q_d \, \ol d_1 \gamma_\mu d_2
.\end{eqnarray}
Here 
$q_u$,
$q_d$ 
are the light quark electric charges. 
The current insertion effectively injects momentum by 
changing flavors,  from 
$u_1$ 
to 
$u_2$ 
in the case of 
$J_\mu^1$, for example. 
These quark-level operators must of course be matched onto the effective theory.

At leading order the matching has already been performed
in Eq.~\eqref{eq:L} as we have minimally coupled the  
source field $A^a_\mu$. 
As explained in~\cite{Tiburzi:2005hg},
the generators $\ol T {}^a$ can be chosen purely in the valence 
sector. The simplest choice is the obvious one:
$(\ol T {}^1)_{ij} = \delta_{i 2} \delta_{j 1} $ 
corresponding to the current 
$J_\mu^1$, 
and 
$(\ol T {}^2)_{ij} = \delta_{i 3} \delta_{j 4}$
corresponding to
$J_\mu^2$. 
At next-to-leading order there is only one additional term
needed in the effective theory.%
\footnote{
There are additional couplings that are exactly cancelled
by wavefunction renormalization. Not surprisingly, 
these can be removed 
from the current by a field redefinition. 
} 
This local correction to the current has the form
\begin{equation}
\d J^a_\mu
= 
i \a_9 
\tilde{D}_\nu
\str 
\Big[
\ol T {}^a
\left(
[ \tilde{D}_\mu \Sigma^\dagger, \tilde{D}_\nu \Sigma ] 
- 
[ \tilde{D}_\nu \Sigma^\dagger, \tilde{D}_\mu \Sigma ] 
\right)
\Big]
.\end{equation}
This is just a rewriting of the current derived from the relevant term
of the Gasser-Leutwyler Lagrangian~\cite{Gasser:1983yg}. 
As 
$\d J^a_\mu$ 
will only contribute to tree level, it will 
not appear in the expression for the finite volume correction. 
It is instructive to note that this current depends only on 
the flavor non-singlet part of the generators. 
Any local terms in the effective theory
arising from flavor singlet terms
will be inaccessible using TwBCs,
as these must appear 
$\propto \str ( \ol T {}^a ) = 0$.

Having written down the matching of the current
in the effective theory up to next-to-leading order, 
we can now deduce the pion form factor at finite
volume. 
As mentioned above, we work in the $p$-regime~\cite{Gasser:1987zq}
with time treated as infinite in extent.
Furthermore, we choose to calculate the form 
factor from the time-component of the current
as is done in the actual lattice calculations, e.g.~\cite{Simula:2007fa,Boyle:2008yd}. 
The loop diagrams contributing to the form factor are depicted in~\cite{Jiang:2006gna}. 
The hairpin contributions exactly cancel, which is not surprising given 
that flavor neutral mesons are electrically neutral. 
As a check on our calculation, we have verified
that the infinite volume form factor agrees
with the known $SU(4|2)$ partially quenched expression%
~\cite{Arndt:2003ww,Bunton:2006va}.
Furthermore when $m_j = m_u$, the partially quenched
expression turns into the well-known infinite volume
$SU(2)$ result%
~\cite{Gasser:1984ux}. 
The finite volume modification 
$\d \cM (L) \equiv \cM(L) - \cM(L = \infty)$ 
to the time-component of the current matrix element is given by
\begin{eqnarray}
&& \d \cM(L)
=
2 E_\pi(\theta) 
\frac{1}{f^2} 
\int_0^1 dx 
\Bigg[
-
 I_{1/2} \left( \frac{\bm{\theta}}{L}, m_{ju}^2 \right)
\phantom{spacers}
\notag \\
&& \phantom{space} +
 I_{1/2} \left(  ( 1 - 2 x) \frac{\bm{\theta}}{L} , \, m_{ju}^2 + 4 x (1-x) \frac{\bm{\theta}^2}{  L^2} \right)
\Bigg]
,\end{eqnarray}
where we have set the overall charge 
$q_u - q_d = 1$. 
The pion energy is 
$E_\pi(\theta) = \sqrt{m_\pi^2 + \bm{\theta}^2 / L^2}$,
and the finite volume modification is encoded in the function 
$I_{1/2}(\bm{v}, m^2)$~\cite{Jiang:2006gna}. 
There is no isospin breaking term in the above expression
as it would have to be proportional to the difference of
the initial and final state energies. 
Additionally there is a symmetry under $\theta \to - \theta$. 
Both of these desirable features are characteristics of the 
Breit frame kinematics.

\begin{figure}
\epsfig{file=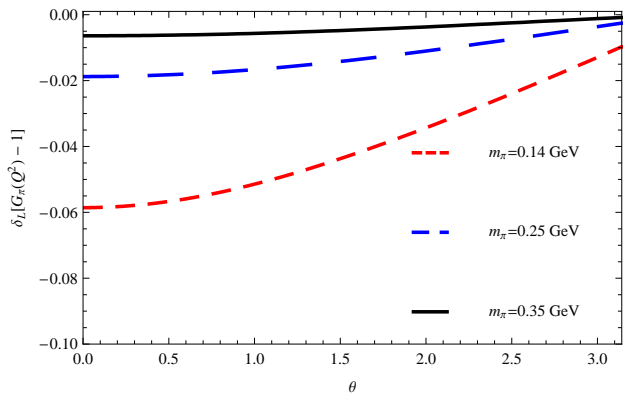,width=8.5cm}
\caption{
Finite volume shift of the pion form factor in the Breit frame.
Plotted vs.%
~$\theta$ 
is the relative change 
$\d_L [ G_{\pi} ( Q^2) - 1]$,
where $Q^2 = \theta^2 \times 0.025  \, \texttt{GeV} {}^2$. 
}
\label{f:FV}
\end{figure}

Numerically we can estimate the finite volume effect. 
For ease we consider the unitary point, $m_{ju}^2 = m_\pi^2$, 
and choose $\bm{\theta}$ to be non-vanishing in a single lattice direction. 
We consider the relative difference 
\begin{equation}
\d_L [ G_\pi (Q^2) - 1 ]
= 
\frac{ \d \cM(L) / 2 E_\pi(\theta) }{G_\pi(Q^2) - 1}
,\end{equation}
where $Q^2 = 4 \theta^2 / L^2$ is the momentum transfer squared,
and the form factor $G_\pi(Q^2)$ has the usual definition:
$\langle \pi(- \theta) | J_4 | \pi(\theta) \rangle = 2 E_\pi(\theta) G_\pi(Q^2)$. 
In Fig.~\ref{f:FV}, 
we hold the lattice size $L$ fixed at $2.5 \, \texttt{fm}$, 
and plot the relative difference 
$\d_L [ G_\pi (Q^2) - 1 ]$
as a function of 
$\theta$
for a few values of the pion mass. 
On current lattices, the volume corrections 
are at the percent level and become 
non-negligible only for light pions 
with small twists.

%
%
%
%
%
%
%


\noindent
{\bf Conclusion}.
Above we investigate an extension of \PQCPT\  
which is relevant  for lattice calculations using 
TwBCs in the Breit frame. 
For hadrons consisting of two light quark flavors, 
the appropriate theory has an enlarged $SU(6|4)$
flavor group. 
The additional valence quarks are fictitious flavors
differing only in their boundary conditions. 
This theory is described in~\cite{Boyle:2007wg}, and
we utilize it here to determine finite volume corrections
for lattice calculations of the pion electromagnetic form factor. 
The result we derive is quite compact, 
maintains isospin symmetry, 
as well as a discrete rotational symmetry.
This is in contrast to the result derived 
employing rest frame kinematics, 
which possessed both 
isospin breaking and cubic symmetry 
breaking terms from finite volume effects~\cite{Jiang:2006gna}.
While numerically the finite volume 
effect is demonstrated to be small for either
kinematics, it would be interesting to 
investigate nucleon isovector quantities
in the Breit frame. 
Results for the nucleon isovector magnetic moment
suggested a larger than expected volume
correction arising from cubic 
symmetry breaking terms~\cite{Tiburzi:2006px}.
The Breit frame kinematics
could mitigate such effects.


\begin{acknowledgments}
We thank S.~Simula for correspondence.
This work is supported in part by the 
U.S.~Dept.~of Energy,
Grant No.~DE-FG02-93ER-40762
(B.C.T.),
and by the 
Schweizerischer Nationalfonds
(F.-J.J.).
\end{acknowledgments}


\bibliography{hb}


\end{document}